# Direct observation of melting in a 2-D superconducting vortex lattice

I. Guillamón,[1*] H. Suderow,[1] A. Fernández-Pacheco,[2,3,4] J. Sesé,[2,4]

R. Córdoba,[2,4] J.M. De Teresa,[3,4] M. R. Ibarra,[2,3,4] S. Vieira [1]

[1]*Laboratorio de Bajas Temperaturas, Departamento de Física de la Materia Condensada, Instituto de Ciencia de Materiales Nicolás Cabrera, Facultad de Ciencias, Universidad Autónoma de Madrid, E-28049 Madrid, Spain*

[2]*Instituto de Nanociencia de Aragón, Universidad de Zaragoza, Zaragoza, 50009, Spain*

[3]*Instituto de Ciencia de Materiales de Aragón, Universidad de Zaragoza-CSIC, Facultad de Ciencias, Zaragoza, 50009, Spain*

[4]*Departamento de Física de la Materia Condensada, Universidad de Zaragoza, 50009 Zaragoza, Spain*

**Topological defects such as dislocations and disclinations are predicted to determine the two-dimensional (2-D) melting transition[1–3]. In 2-D superconducting vortex lattices, macroscopic measurements evidence melting close to the transition to the normal state. However, the direct observation at the scale of individual vortices of the melting sequence has never been performed. Here we provide step by step imaging through scanning tunneling spectroscopy of a 2-D system of vortices up to the melting transition in a focused-ion-beam nanodeposited W-based superconducting thin film. We show directly the transition into an isotropic liquid below the superconducting critical temperature. Before that, we find a hexatic phase,**

**characterized by the appearance of free dislocations, and a smectic-like phase, possibly originated through partial disclination unbinding. These results represent a significant step in the understanding of melting of 2-D systems, with impact across several research fields, such as liquid crystal molecules, or lipids in membranes [4–7].**

The seminal ideas of Kosterlitz and Thouless [1], further developed by Halperin, Nelson and Young [2,3], have been much discussed in connection with melting of many different 2-D hexagonal crystals. These include crystals formed from elongated rod-like entities such as superfluid vortices, lipids or liquid crystal molecules, as well as from more isotropic constituents, as electrons Wigner crystals, particles or bubbles, or electronic charge density arrangements[8–10]. Theoretically, it is found that the unbinding of dislocation pairs creates an intermediate phase in a 2-D hexagonal crystal, named the hexatic phase, which retains sixfold orientational order but has no long range translational order[1–3]. The hexatic phase, which appears between the 2-D crystalline phase and the isotropic liquid, has been since then subject of much research. Vortex lattices in superconducting thin films appear as an ideal system where the mechanism of the 2-D melting process can be investigated. Until now, work on the 2-D solid-liquid superconducting vortex lattice transition has focused on macroscopic studies of thermal properties or of critical currents around the melting transition [11–15]. The transition seems to be consistent with the theory of 2-D dislocation-unbinding melting behavior[1–3, 16–18]. However, no local direct observation of melting has been provided. Here we track the modifications induced by temperature on a 2-D vortex system, from its formation up to the isotropic liquid state. Increasing temperature we first observe thermal de-pinning of vortices, which produces a more ordered hexagonal lattice. Melting process is then initiated with the appearance of free dislocations, corresponding to the hexatic phase, from which a smectic-like phase emerges, before the isotropic liquid is formed.

Our sample is an amorphous W-based thin film nano-deposited using focused-ion-beam, with a superconducting critical temperature at zero field $T_c$ = 4.15 K[19,20]. It is an extreme type II superconductor, with a coherence length $\xi$ = 6.25 nm and a London penetration depth $\lambda$ = 850 nm (see supplementary information). The sample's thickness (200 nm) is far smaller than $\lambda$. Therefore, vortices behave to a large extent as straight parallel lines[11], forming a 2-D solid, as in other extreme type II amorphous thin films[17,21], where 2-D melting has been studied through macroscopic measurements.

The superconducting vortex lattice can be imaged using Scanning Tunneling Microscopy and Spectroscopy STM/S[22]. As the superconducting gap closes inside the vortex cores, by means of tracking the Fermi level (zero bias) tunneling conductance, normalized to the value at voltages well above the superconducting gap, as a function of the position at the sample's surface $\sigma(x, y)$ well resolved vortex lattice (STS) images can be obtained. However, very stable and flat surfaces are required, with homogeneous superconducting tunneling conductance curves at zero field. Previous work in extreme type II superconducting thin films needed to evaporate an additional small normal surface Au layer[23,24]. This strongly decreased the contrast and imaging capacities of the STM/S, so the vortex lattice has not been followed as a function of temperature. Other local real-space techniques have faced different resolution limitations close to melting [25,26].

By contrast, in the thin films studied here, STM/S can be used in a wide range of temperatures and magnetic fields without a particular surface preparation[20]. Topographical (STM) landscapes are highly uniform. Large flat areas are easily found, whose main accidents are small and smooth hills, separated by linear surface features of a few nanometers height and around two hundred nanometer lateral size. At zero field, tunneling characteristics (STS) show the features expected for a well behaved BCS superconductor, with a uniform gap value $\Delta = 0.66$ meV over the whole surface[20]. Due to the flatness of the sample surface and the uniformity of its superconducting behavior as probed by STM/S, we have been able to follow with detail the temperature induced changes in small groups of vortices. Contrast is maximized in STS images, so that inter-vortex superconducting state is represented as white, and the normal intra-vortex state as black[20]. Besides information about vortex positions, $\sigma(x,y)$ also shows directly if the sample is superconducting or normal (where $\sigma(x,y) = 1$), which is crucial to identify the isotropic liquid. Vortex images are made in a fixed location and at constant temperature and magnetic field (applied at low temperatures in zero field cooled conditions). Temperature was increased in steps of 0.1 K, and each image was taken as fast as possible, in most cases in about 8 minutes. Several images were taken at a single temperature. More detailed information is given in the supplementary information. At low temperatures, typically we observe a triangular lattice which is rather disordered due to pinning by features of the topography of the surface[20]. Images have been taken at up to ten different locations, and at four magnetic fields (1 T, 2 T, 3 T and 4 T), always finding the phenomenology described by the individual regions shown below, so that the observations are representative for the whole surface.

Thermally induced de-pinning is shown in Fig. 1, with several representative vortex lattice images. Here we have searched for an area showing several surface features which act as pinning centers. When increasing temperature, the distorted vortex lattice does not change up to 1.5 K. Vortices move and re-arrange in the temperature interval between 1.5 K and 1.6 K. De-pinning occurs essentially at the central part of the image, and leads to a more ordered vortex lattice. STS images and the corresponding video (see supplementary information) are self-explanatory. Nevertheless, it is convenient to discuss the temperature evolution of their Fourier transforms. These show the six Bragg peaks associated to the hexagonal symmetry of the vortex lattice with an angular distortion at low temperatures, which disappears abruptly at the de-pinning temperature of 1.5 K (Fig. 1**f**). So the lattice disorder induced through vortex pinning by features of the surface topography, decreases well below the appearance of the melting regime.

Let us now focus on the sequence of results approaching $T_c$, disclosing the melting process in detail. The starting point in these experiments are hexagonal vortex bundles produced by thermal de-pinning, with typically two to three hundred nanometer lateral size. Moreover, we select flat surfaces to avoid, as far as possible, the effect of vortex pinning. Fourier transform of the images shows six-fold Bragg peaks at $2\pi/d$ (as in upper right inset of Fig. 1**f**; $d$ is the lattice constant). The average between the Fourier amplitude of the six peaks decreases continuously with temperature (Fig. 2, at H = 2 T). When the amplitude of the Bragg peaks becomes zero, we observe that the superconducting signal is not lost in the tunneling conductance curves. $\sigma(x, y)$ is below one and uniform over the whole surface, evidencing a homogeneous open superconducting gap. So $\sigma(x, y)$ averaged over the whole image $\sigma_{average}$ (Fig. 2) increases continuously through the point where the Bragg peak amplitude goes to zero. At this point the isotropic vortex liquid forms, and vortices move below the tip, due to thermal excitation, much faster than the imaging time, so that $\sigma(x,y)$ obtained is the average between result in- and outside vortices. Finally, at 3.2 K, the full disappearance of superconducting features in the tunneling conductance curves, with $\sigma(x,y)$ equal to one and showing no longer any temperature dependence, marks the transition to the normal state.

In Fig. 3 we show six representative STS images of the melting process at H = 2 T. At 1.2 K the vortex lattice shows an ordered hexagonal arrangement in most of the image (Fig. 3**a**). When increasing temperature, around 1.9 K, we begin to observe isolated dislocations in the previously ordered lattice (Fig. 3**b**). They remain at the same position at a fixed temperature, but they move when the temperature is changed (Fig. 3**b-d**). We consider the appearance of free dislocations as the start of a hexatic phase. At slightly higher temperature, vortices of the upper part of the images start to be blurred forming a partially striped image (Fig. 3**d**). Vortices have a strong 1-D temperature induced motion along curved stripes, having the low temperature lattice constant $d$ as the inter-stripe distance. A new smectic-like vortex phase can be identified by the presence of these curved stripes. In the present case (Fig. 3), a disclination may have been formed with the center close to the scanned area. So the formation of the smectic phase can be associated to partial disclination unbinding. At 2.9 K, the superconducting gap is observed over the whole image, but there are no isolated vortices nor stripes over large parts of the image, evidencing a fully isotropic liquid, as shown at 3 K in Fig. 3**f**. Images show homogeneous superconducting tunneling features, with vortices freely moving due to thermal excitation. At even higher temperatures, in the normal phase, images are homogeneous and without any superconducting signal. We have followed similar images at very different densities, with lattice constants ranging from $d = 50$ nm at 1 T to $d = 25$ nm at 4 T, and in up to ten different places on the surface, finding always similar behavior. Appearance of free dislocations occurs around 1 K, and the isotropic liquid is formed around 0.3 K, below the transition to the normal phase (at $H_{c2}$(T), which is shown in the supplementary information).

We often observe coexistence of liquid with hexatic, as well as with smectic-like regions. Clearly, the transition happens for bundles of vortices. Although it appears very difficult to make quantitative statements from our images showing only a small amount of vortices, the disorder produced by the the small surface roughness may play a role in the size of the bundles, and possibly have some influence, which is yet to be worked out, in the observed phenomena. On the other hand, with the appearance of smectic-like and isotropic liquid phases, the amplitude of thermal vortex motion is strong enough to produce blurred structures. Remarkably, these two phases mark the step-like loss of orientational order, which is maintained in the hexatic phase, where thermal vortex motion is not strong enough to blur isolated vortices in our STS images.

The emerging picture from our data is that, in addition to the appearance of free dislocations originating the hexatic phase, there is another Kosterlitz-Thouless type transition from enhanced vortex mobility along characteristic curved lines formed by disclinations, which produces a smectic-like phase. The observed three stage process could be a general mechanism of melting in 2-D, at least for solids formed by linear-like units, as large molecules in liquid crystals and vortices in Abrikosov lattices. It would be interesting to know if this conclusion can be extended to other superfluids where 2-D vortex lattice conditions can be created, as e.g. thin liquid helium films or atoms in optical lattices.

In summary, we have performed a direct observation of several phenomena induced by temperature in a 2-D lattice of vortices. Thermal de-pinning from surface irregularities was observed, as well the formation of an isotropic vortex liquid. In between these, two Kosterlitz-Thouless type transitions, characterized by the appearance of free dislocations and of stripes, govern the melting process. We believe that our findings should be relevant to understand the behavior of vortices in other superconductors, and in a more general context for the physics of melting in 2-D systems.

**Acknowledgements** We acknowledge discussions with F. Guinea and A.I. Buzdin. The Laboratorio de Bajas Temperaturas is associated to the ICMM of the CSIC. This work was supported by the Spanish MICINN (Consolider Ingenio Molecular Nanoscience CSD2007-00010 program, MAT2008-06567-C02 and FIS2008-00454), by the Comunidad de Madrid through program "Science and Technology at Millikelvin", by the Aragon Regional Governement, and by NES and ECOM programs of the ESF.

**Competing Interests** The authors declare that they have no competing financial interests.

**Correspondence** Correspondence and requests for materials should be addressed to I.G. (email: isabel.guillamon@uam.es).

**Figure 1   Thermal de-pinning of the 2-D vortex lattice.** (**a**) STM topography, with grey scale corresponding to height changes of 10 nm. Surface topography features are marked in **a**-**e** as red dashed lines. (**b**) STS vortex lattice image after applying a magnetic field of 1 T at 0.1 K. Vortices (black areas) form a hexagonal lattice, which is strongly distorted, as highlighted by the light blue hexagon. Note that vortices have a strong tendency to follow a near square arrangement, imposed by the linear surface features surrounding the central part of the image. When increasing temperature, nothing changes until we reach 1.5 K (**c**-**d**), where vortices start to move. The changes are subtle. The vortex lattice in the central part of the image suffers a small rotation, which we highlight by plotting magenta and blue hexagons. Magenta hexagon shows present vortex arrangement, whereas blue hexagon shows the same six vortices at somewhat lower temperatures. The vortex arrangement switches (as schematically shown by orange arrows) between two positions, although the temperature is fixed. Above 1.6 K (**e**), the vortex lattice has found a stable position, with a more ordered triangular arrangement. (**f**) Angular distortion of the vortex lattice as measured from an analysis of the Fourier transform of the vortex lattice images. The angles $\alpha_1$ and $\alpha_2$, as defined in the images, which represent the Fourier transform of the STS images at 0.1 K -lower left- and 1.6 K -upper right-, strongly change at the de-pinning temperature.

**Figure 2   Temperature induced changes in the superconducting signal and in the contrast of the vortex lattice STS images.** Amplitude of the six-fold Bragg peaks (open circles) of the Fourier transform of images, and the normalized zero voltage tunneling

conductance averaged over whole images $\sigma_{average}$ (closed circles), as a function of temperature and at 2 T. $\sigma_{average}(T)$ gives the decrease in the superconducting signal in the STS images due to temperature induced smearing of the local density of states and decrease of the superconducting gap (see also [20]). The Fourier amplitude of the Bragg peaks (open circles) shows the decrease in vortex lattice contrast in the STS images. This has a steeper temperature dependence than $\sigma_{average}$ because of thermal motion of vortices. Time scale of thermal fluctuations in the vortex positions is more than ten orders of magnitude below the time we need to make a line scan over each vortex (some ten s)[11]. Therefore, the obtained tunneling conductance curves give the average between those inside and outside the vortices, which continuously move below the tip. Vortex lattice contrast is lost well below $T_c$, signalling the transition to an isotropic liquid.

**Figure 3  Three stage melting.** (**a**) Hexagonal lattice as imaged by STS at a magnetic field of 2 T (lateral size is of 220 nm, individual vortices are the black areas). Image is taken above the de-pinning temperature, so the vortex lattice is more ordered than at lower temperatures. Magenta dashed lines highlight the vortex arrangement. At 1.9 K (**b**), we start to see temperature induced changes. The lattice shows more disorder, and a dislocation enters the image from its bottom left part. This marks the hexatic phase. Magenta lines highlight the dislocation, and the orange and green lines corresponding seven-fold and five-fold coordinated vortices. Slight temperature increase (**C**) further moves the dislocation to the upper part of the figure. At 2.1 K (**d**), a new feature appears in the images. Upper right part of the image (**d**) starts showing linear vortex arrangements (highlighted

by dashed magenta lines). Local contrast is lost along these lines, evidencing strong one dimensional motion. Note that movement occurs along curved paths very similar to those produced by disclinations. Moreover, separation between lines corresponds to the low temperature inter-vortex distance. This evidences the presence, at the local level, of a smectic-like phase. The smectic-like phase coexists with the disordered hexatic vortex lattice seen at the bottom of the image. (**e**) At the lower right part of the image, no stripes nor isolated vortices are seen. Nevertheless, smectic-like phase is still roughly visible at the upper part. (**f**) Finally, no isolated vortices nor smectic-like arrangements are seen, and the superconducting gap is homogeneous, evidencing that the vortex lattice has formed an isotropic liquid.

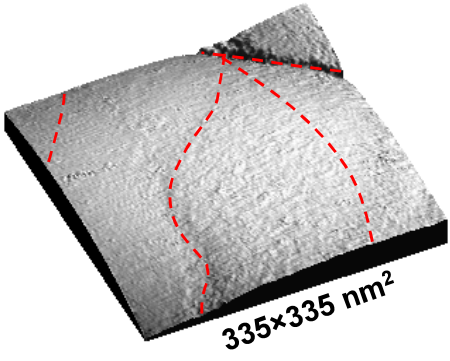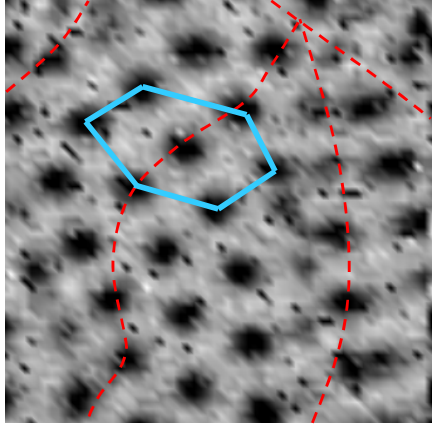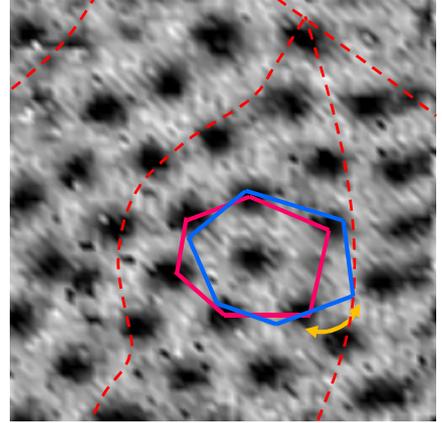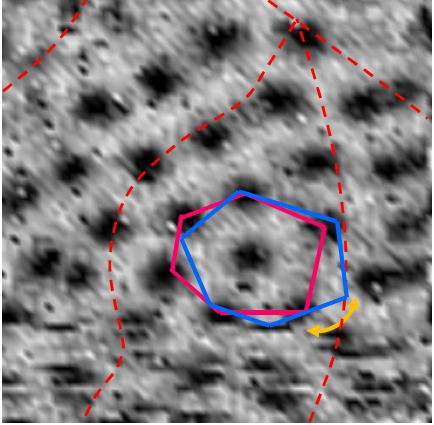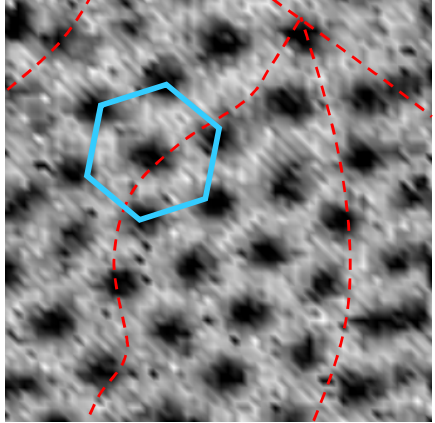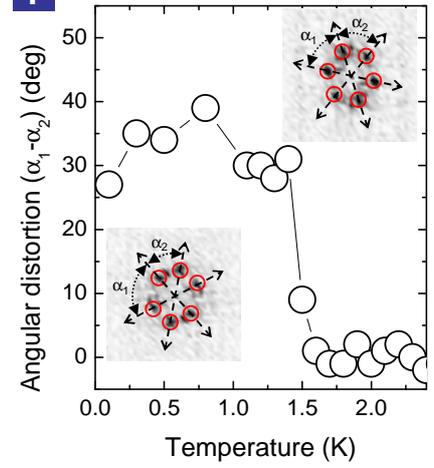

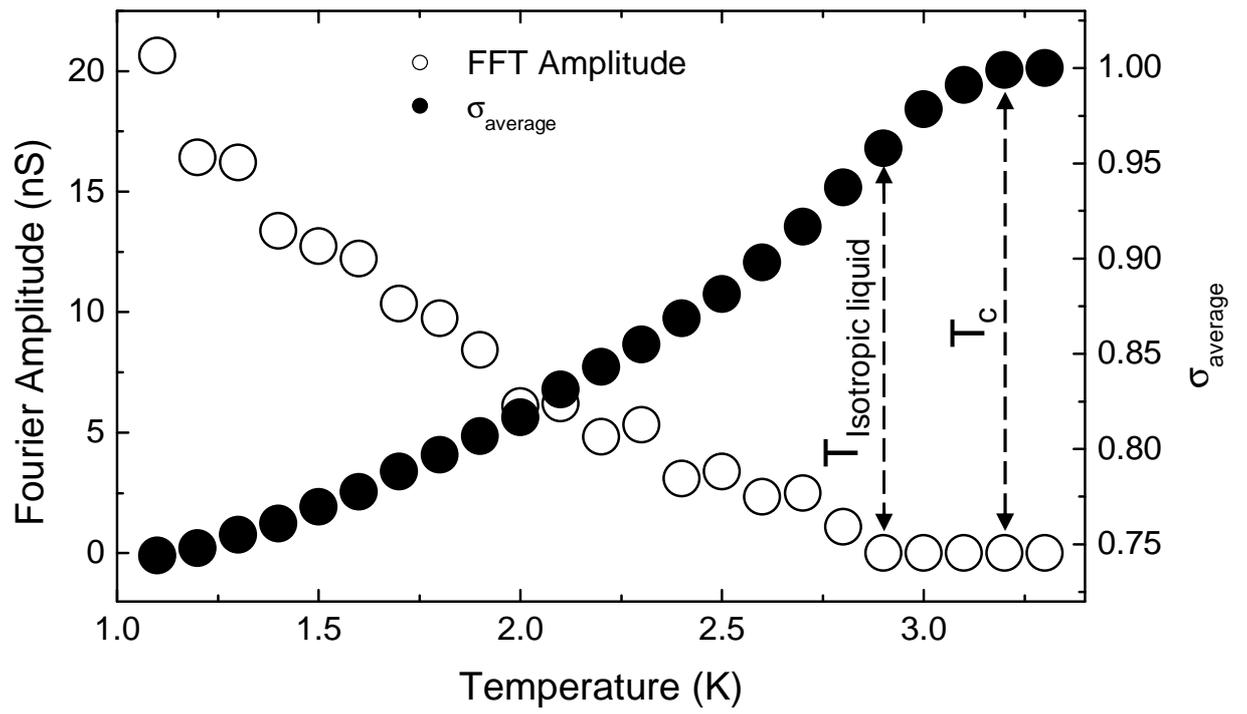

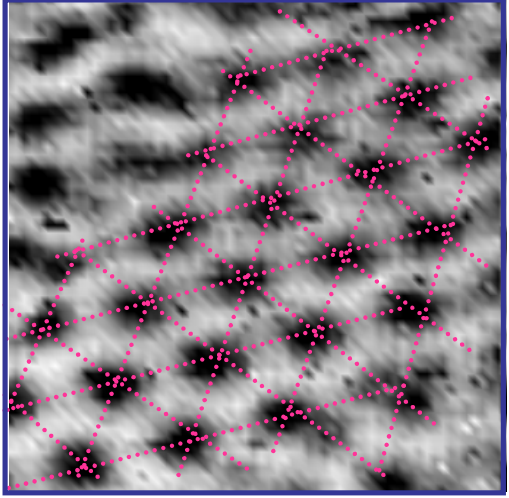
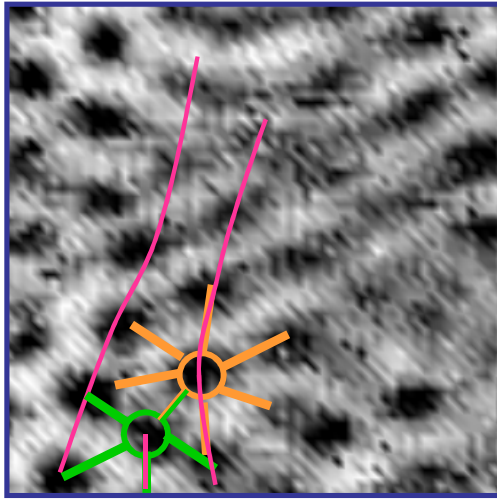
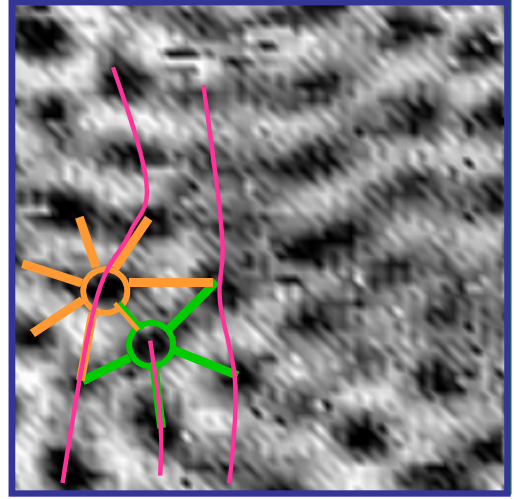
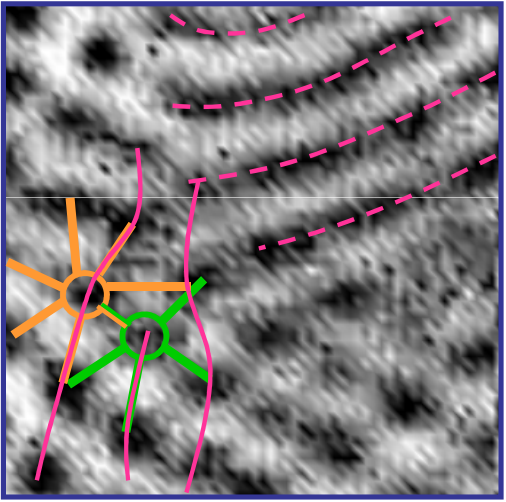
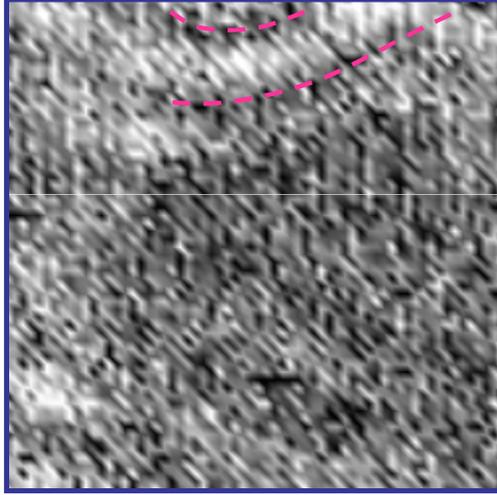
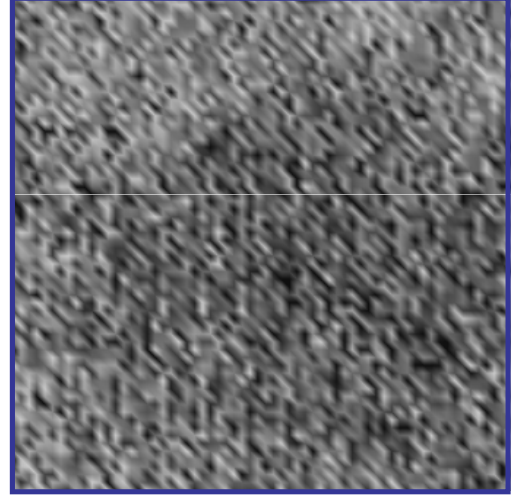

# SUPPLEMENTARY INFORMATION

# Direct observation of melting in a 2-D superconducting vortex lattice

I. Guillamón,[1] H. Suderow,[1] S. Vieira,[1] A. Fernández-Pacheco,[2,3,4]
J. Sesé,[2,4] R. Córdoba,[2,4] J.M. De Teresa,[3,4] and M.R. Ibarra[2,3,4]

[1]Laboratorio de Bajas Temperaturas, Departamento de Física de la Materia Condensada
Instituto de Ciencia de Materiales Nicolás Cabrera, Facultad de Ciencias
Universidad Autónoma de Madrid, E-28049 Madrid, Spain
[2]Instituto de Nanociencia de Aragón, Universidad de Zaragoza, Zaragoza, 50009, Spain
[3]Instituto de Ciencia de Materiales de Aragón, Universidad de Zaragoza-CSIC, Facultad de Ciencias, Zaragoza, 50009, Spain
[4]Departamento de Física de la Materia Condensada,
Universidad de Zaragoza, 50009 Zaragoza, Spain

## DESCRIPTION OF STM/S SET-UP AND OF IMAGING PROCEDURES

We use a STM/S in a partially home-made dilution refrigeration system with a 9 T magnet. The fully home-made STM head and electronics allow for scanning windows up to $2 \times 2 \mu m^2$, and has been previously used to image the vortex lattice in different systems (S1-S3). The 200 nm thick W deposits for the STM study were grown by sweeping the focused ion beam onto $30 \times 30 \mu m^2$ areas on top of a 160 nm-thick Au-Pd layer previously deposited by evaporation on a Si substrate, as we describe in Ref.[20] of the publication. The tip was positioned on top of the W deposit at room temperature, and the whole STM/S set-up was subsequently cooled. As shown in Ref.[20] of the publication, the tunneling conditions were excellent, with high valued work functions (several eV), and fully reproducible superconducting tunneling conductance curves.

Vortex lattice STS images are built from the normalized zero bias conductance changes as a function of the position $\sigma(x,y)$ at fixed temperature and magnetic field, as in Refs.(S1-S3) and Ref.[20] of the publication (see also Refs.(S4-S9) for work from other groups). We take a full I-V curve at each position and make a numerical derivative which gives us the zero bias conductance. Here, we have considerably improved the imaging speed to be able to take a single image in eight minutes. This allows us to study temperature and field induced variations with enough density of points and to identify slow dynamic processes occurring at a fixed temperature. We check that eventual vortex lattice changes have occurred before taking the image, usually in time scales far smaller than the imaging time.

Dilution refrigerator naturally provides for an extremely well controlled temperature environment, which we can vary in steps between 0.1 K and 6 K with a PID control (Air Liquide TRMC2) that provides for a stability of better than 0.1 %. We use a Germanium thermometer calibrated by Lakeshore, and a Matshushita carbon thermometer located in the zero field region of the magnet to avoid for magnetoresistance effects. Red dashed lines in figures and videos of the vortex lattice represent the position of main features of the surface topography, as seen in the simultaneously acquired STM images.

No filtering nor any other image treatments have been used. In the microscopy STM images, we subtract a plane as usual, and we adjust the grey scale for showing most clearly height differences, which corresponds to 10 nm or less. In the STS images, we adjust the grey scale at each image to best view the position of individual vortices. Actually, the gray scale of the STS images is adjusted in such a way that the difference between black and white closely follows the difference between one and $\sigma_{average}(T)$ shown in Fig. 2 of the publication.

Thermal drift of the STM/S set-up is completely negligible in this temperature range, as evidenced by simultaneously taken topography STM images. However, note that even atomic-scale topographic studies in other samples (of NbSe$_2$) show that, up to 5 K, the scanning window does not move when increasing temperature. This is not surprising, as the thermal expansion of the materials used in the STM/S set-up becomes negligible in this temperature range.

## COMMENT ON VORTEX LATTICE MELTING IN HIGH T$_c$ MATERIALS.

Note that vortex lattice melting in High T$_c$ materials has been discussed to a large extent (see Ref.[12] of the publication). However, observation of single vortices remains a formidable task, and the vortex lattice cannot be considered a simple 2-D solid, as vortices may decouple between planes. This has been shown by Oral et al.(S10), who viewed the vortex lattice melting by increasing the magnetic field at fixed temperature and magnetic fields of several tenths of mT. To the best of our knowledge, there are no experiments showing thermally induced melting tracked through individual vortex observations in these materials. Other systems where vortex lattice melting has been considered are Bose-Einstein condensates (see e.g. *(S11)*), still without corresponding experimental observations (to our knowledge).

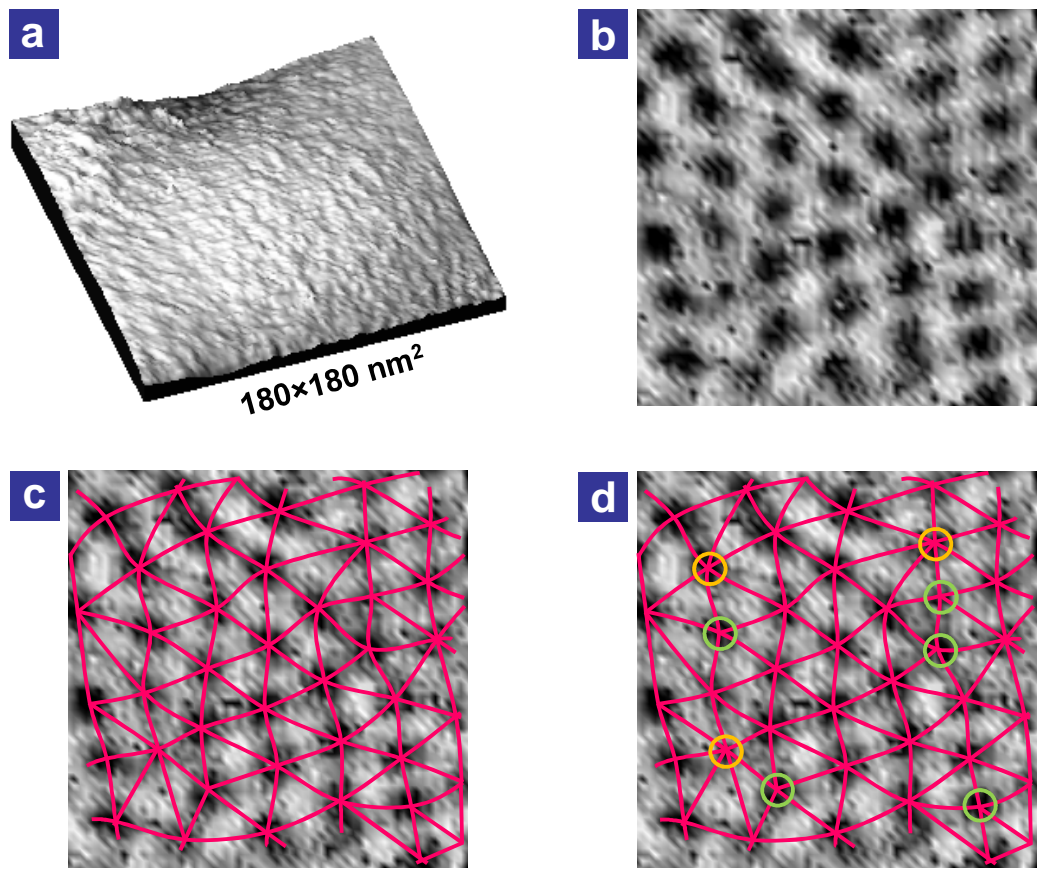

**Figure S** 1: **Hexatic phase.** Vortex lattice image with a large amount of dislocations within the hexatic phase (at 1.6 K and 3 T). (**a**) Corresponding STM topography. Note that this area is particularly flat, with a total height change below 2 nm. (**b**) Bare vortex lattice STS image. (**c**) Same image with triangulated vortex positions. (**d**) five-fold (green) and seven-fold (orange) coordinated vortices found in the image. Most of them pair, and form dislocations, although there are two five-fold coordinated vortices, close to the right and right-bottom edges of the image, whose seven-fold partner cannot be identified clearly within the scanned area.

### SUPPLEMENTARY FIGURES.

In Fig.S1 we show the vortex arrangements above the melting transition, in the hexatic phase. The melting phase diagram with the temperature and field regions for the appearance of hexatic, smectic and isotropic liquid phases is shown as an inset in Fig.S2. The Fig.S2 shows the whole temperature dependence of $H_{c2}(T)$ found in this thin film. In Fig.S3 we show, for completeness, the topography of the area discussed in Fig.3 of the publication.

### BRIEF DESCRIPTION OF SUPPLEMENTARY VIDEOS.

*Supplementary Video 1: depinning.wmv* A set of 90 subsequent STS images when stepwise increasing the temperature from 0.1 K to 2.1 K at a fixed field of 1 T. The features of the simultaneously acquired topography STM images are highlighted by red dashed lines and the vortex arrangements by blue and magenta hexagons. At each image, the corresponding temperature is shown in the top. Around the de-pinning temperature, we highlight arrangements

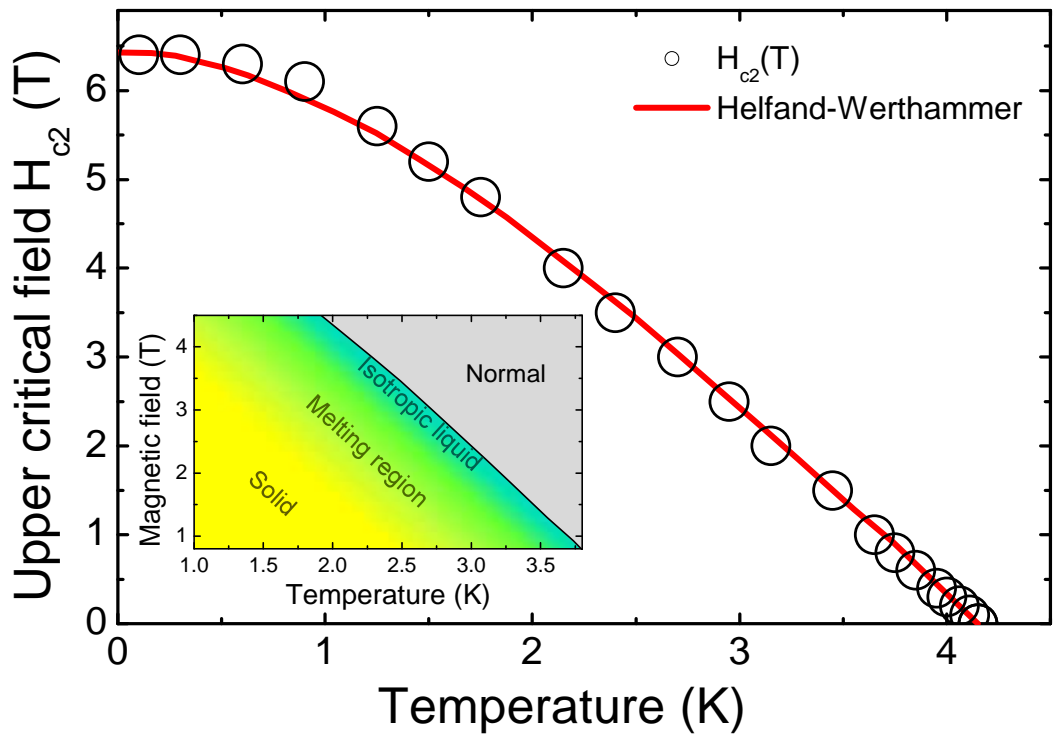

**Figure S 2**: **Upper critical field $H_{c2}(T)$ and melting phase diagram.** The upper critical field $H_{c2}(T)$ has been followed closely as a function of temperature by recording the position in the H-T diagram where we lose all superconducting features in the tunneling spectroscopy curves. The line is given by Helfand-Werthammer theory of $H_{c2}(T)$ in the dirty limit *(S12)*. Inset shows schematically the vortex phase diagram which can be extracted from our measurements. Using usual procedures in thin films of amorphous superconductors (see e.g. Ref.[21] of the publication), we extract a Ginzburg-Landau parameter $\kappa = 83$, a London penetration depth $\lambda = 850 nm$ and a superconducting coherence length $\xi = 6.25 nm$ from $H_{c2}(T)$ and the residual resistivity of our sample ($\rho_0 = 275 \mu\Omega cm$). All these parameters are similar to those previously discussed in many amorphous superconducting metals made of transition metal elements. They allow us to estimate the relative width of the liquid regime through the Levanyuk-Ginzburg number *(S13-S14)*, $G_i = 1/2[k_B T_c/4\pi\xi^3(0)\mu_0 H_c^2(0)]^2$ (here $k_B$ is Boltzmann's constant, and $H_c$ the thermodynamic critical field, see Refs.[11,12] of the publication). We find $G_i = 1.15 \cdot 10^{-4}$, and obtain that the formation of an isotropic liquid should appear around 300 mK below $H_{c2}(T)$, as actually discussed in the publication. Let us note that, with such a large Levanyuk-Ginzburg number, the vortex lattice melts well before $\xi(T)$ increases in such a way as to overlap vortex cores and eventually reduce contrast in STS images. As shown in our images, before the vortex lattice forms an isotropic liquid, isolated vortices do not change its size. The strong upturn in $\xi(T)$ becomes significant only some tens of mK close to $T_c$.

showing hexagon with the same convention as in the publication. Namely, magenta hexagon is the arrangement at the present temperature and blue hexagon the previous arrangement at a somewhat lower temperature. Each image takes about 8 minutes, and we spend about one hour at each temperature, so that several images are shown at fixed temperature in the video. This allows to clearly identify the switching between two arrangements observed at 1.5 K and at 1.6 K.

*Supplementary Video 2: melting.wmv* A set of 68 vortex lattice images taken at 2 T when increasing temperature from 1.1 K up to 3.2 K at 2 T. Features of vortex arrangements are highlighted by magenta lines, as in Fig. 3 of the publication. Dislocations and stripes are also highlighted as in Fig. 3 of the publication. At each image, the corresponding temperature is shown in the top.

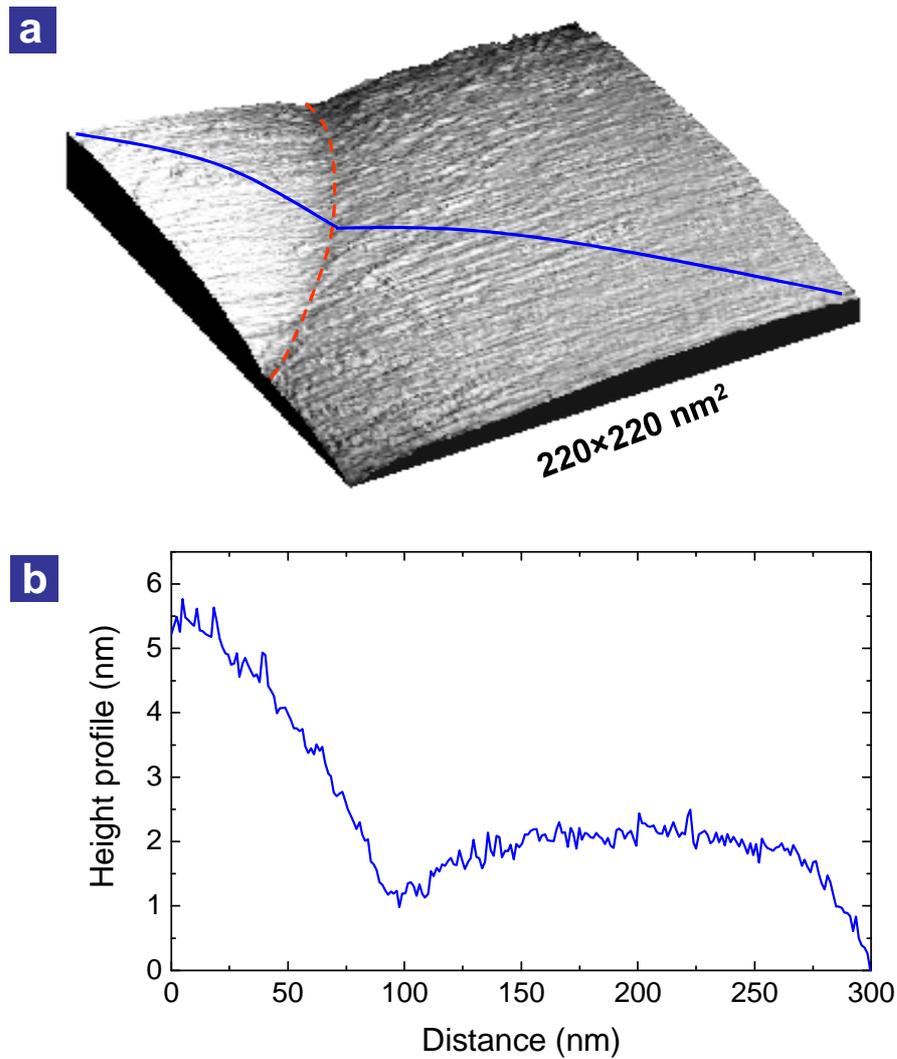

**Figure S** 3: **STM topography of the area discussed in the Fig.3 of the publication.** The topography of this area (**a**) has one visible surface feature (red line) running at the upper left edge of the image. The rest of the image, which also shows a regular hexagonal vortex lattice, is smooth. The blue line in (**a**) is the profile shown in (**b**). Note that the thin film is here particularly flat. The smooth hills observed are only a few nm in height, i.e. some per cent of the total thickness of the superconducting thin film. How these structures are created is not clear. However, their small size points out that they can be easily formed during deposition, or they may be created by the underlying Au-Pd layer (see Ref.[12] of the publication).

# SUPPLEMENTARY REFERENCES